\documentstyle{article}
\def\be{\begin{equation}}
\def\ee{\end{equation}}
\def\bea{\begin{eqnarray}}
\def\eea{\end{eqnarray}}
\def\pa{\partial}

\def\fn{\footnote}

\def\case#1/#2{\textstyle\frac{#1}{#2}}

\begin{document}
\begin{center}
\Large
{\bf Reformulation and Interpretation of the SMS Braneworld}
\vspace{.2in}
\normalsize

{\Large Edward Anderson and Reza Tavakol.}

\vspace{.2 in}

{Astronomy Unit, School of Mathematical Sciences, Queen Mary, University of London \\
Mile End Road, London, E1 4NS, U.K.}

\end{center}

\vspace{0.4in}


\date{}
\begin{abstract}

We reformulate the Shiromizu, Maeda and Sasaki (SMS) braneworlds 
within the framework of the five-dimensional Einstein equations.
In many applications of the braneworld Einstein field equations,
the Weyl term is attributed to the bulk, thus splitting the
non-Einsteinian terms into `bulk' and `brane' terms.
Here by employing standard geometrical identities,
we show that such a split is non--unique, since they get mixed up
in different formulations.
An important consequence of this non-uniqueness is that
even though the full brane-bulk systems in all such
formulations are completely equivalent, important differences can arise were one to
truncate different formulations by throwing away the associated `bulk' terms.
This is particularly likely to be the case in more general 
anisotropic/inhomogeneous settings with non--AdS bulks,
in which the common truncation of the SMS (which throws away the Weyl term) would not coincide
with the full system.
We emphasize that rather than providing support for any particular
truncation, these differences show clearly the dangers
of using {\it any} truncated equations and provide a strong
argument in favour of studying the full brane-bulk system.  
The different formulations we provide also permit different ways of approaching
the full brane-bulk system which may greatly facilitate its study.
An example of this is the second-order nature of
the formulations given here as opposed to SMS's
formulation which is third-order.

\end{abstract}

\hspace{0.08in} PACS number: 04.50+h


\section{Introduction}

Braneworlds are often studied within the framework of the 5-d Einstein 
Field Equations (EFE), a prime example being the general formulation of
Shiromizu, Maeda and Sasaki (SMS) \cite{SMS}.  
According to this, the 
observed universe is confined to a 4-d brane, which is a particular 
timelike hypersurface embedded within a 5-d bulk. 
On the brane, the 10 4-d Einstein field Equations (EFE) are 
replaced by the 10 SMS braneworld EFE (BEFE) \cite{SMS}.
There are two important ways in which these BEFE differ from the EFE. Firstly,
these BEFE do not constitute a closed system,
since they contain an unspecified `electric' Weyl tensor term,
which can only be specified in terms of the 5-d bulk.      
Secondly, they contain a term quadratic in the energy-momentum tensor of the brane,
which can have important consequences for the evolution of very early universe models 
\cite{Bin2,MWBH}.  
This term originates from the junction conditions 
which follow from the assumption that the brane is a thin matter sheet embedded in the bulk, 
about which there is a $Z_2$ reflection symmetry \cite{SMS}.  
Given the complexity of the full SMS brane-bulk system of equations, 
the usual approaches in studies of the braneworld models 
effectively involve throwing away the Weyl term using some 
plausibility arguments, 
or leaving all or part of it as some free unknown.  
However, it is generally accepted that 
in principle none of these approaches 
are entirely satisfactory \cite{Maartensdec}.

Here we reformulate the SMS braneworlds by means of geometrical identities.  
Whereas the non-Einsteinian terms in SMS's BEFE are often interpreted as 
a `bulk' term (the Weyl term) which is then often discarded and `brane' terms to be kept, 
we demonstrate by our reformulations that such a split of the 
non-Einsteinian terms of the BEFE into `bulk' and `brane' parts is non-unique.
Clearly all such reformulations and splits (including SMS's) give completely 
equivalent systems, so long as the full brane-bulk systems is considered.  
Important differences would, however, arise were one to truncate different formulations
by throwing away the corresponding `bulk' parts.
This demonstrates clearly that different 
truncations result in different imprints on the residual `braneworld physics'
and therefore all truncations are unsatisfactory.  
Only the study of the full brane-bulk system is free from such ambiguities.

The different formulations given here also provide 
different ways of studying the full system 
which may be more convenient to study the  
brane-bulk system than SMS's formulation.  
An example of this is that many of our formulations are second-order 
unlike the third-order SMS system.

In Sec 2 we summarize the steps involved in deriving the SMS braneworld EFE. 
In Sec 3 we use a brief account of the origin of the junction conditions 
to demonstrate some of the non-uniqueness in the split of the non-Einsteinian 
terms in the BEFE into 'bulk' and 'brane' terms.     
In Sec 4 we give two formulations of the BEFE with no explicit quadratic terms, 
the second of which has no explicit Weyl term either. 
This demonstrates how completely equivalent formulations of the full
system would give different (in general unsatisfactory) truncated braneworld 
equations were one to remove the corresponding `bulk terms'. 
We conclude in Sec 5.

\section{Braneworld Einstein Field Equations in brief}

The derivation of the SMS braneworld EFE begins with the split of 5-d spacetime w.r.t a 
timelike hypersurface $\Upsilon$ with spacelike normal $n^A$.  
The 5-d metric $g_{AB}$ is thus split according to 
\be
\begin{array}{ll}
g_{AB} = \left( \begin{array}{ll} \mbox{ } \mbox{ } \mbox{ } \mbox{ } \beta_i\beta^i + \alpha^2  &
\beta_b \\ \mbox{ } \mbox{ } \mbox{ } \mbox{ } \beta_a &  h_{ab} \mbox{ } \mbox{ } \mbox{ } \mbox{ } \mbox{ } \end{array}\right)
\mbox{ }  , \mbox{  }  \mbox{  }
\end{array}
\label{ADMs}
\ee
where $ h_{ab}$ is the metric induced on the hypersurface.
The extrinsic curvature of the hypersurface relative to the embedding spacetime is
\be
K_{ab} = -\frac{1}{2\alpha}\delta{h}_{ab},
\label{ecd}
\ee
where the hypersurface derivative is given by
$\delta \equiv \frac{\pa}{\pa z} - \pounds_{\beta}$, with $\pounds_{\beta}$ denoting the Lie
derivative w.r.t $\beta_i$. 
In normal coordinates, $\alpha = 1$ and $\beta_i =0$, so $\delta = \frac{\pa}{\pa z}$.

The three projections of the Riemann tensor are respectively the Gauss, 
Codazzi and Ricci equations 
\be
{\cal R}_{abcd} = R_{abcd} - 2K_{a[c}K_{d]b},
\mbox{ } \mbox{ } \mbox{ } \mbox{ } \mbox{ } \mbox{ } \mbox{ } \mbox{ } \mbox{ } \mbox{ } \mbox{ } \mbox{ } \mbox{ } \mbox{ } \mbox{ } \mbox{ }
\label{Gaussful}
\ee
\be
{\cal R}_{\perp abc} =  -2D_{[c}K_{b]a}, \mbox{ } \mbox{ } \mbox{ } \mbox{ } \mbox{ } \mbox{ } \mbox{ } \mbox{ } \mbox{ }
\mbox{ } \mbox{ } \mbox{ } \mbox{ } \mbox{ } \mbox{ } \mbox{ } \mbox{ } \mbox{ } \mbox{ } \mbox{ } \mbox{ } \mbox{ } \mbox{ } \mbox{ } \mbox{ }
\label{Codazzieq} 
\ee
\be
{\cal R}_{\perp a\perp b} =  \frac{1}{\alpha}(\delta K_{ab} - D_bD_a\alpha) + {K_a}^cK_{cb},
\label{Riccieq}
\ee
where the tensorial objects represented by the capital Roman and Calligraphic 
letters represent the usual quantities in 4-d and 5-d respectively,  
$D_{c}$ denotes the 4-d covariant derivative,
and where we have used  ${\cal O}_{...a...} = {\cal O}_{...A...}h^A_a$ 
for projections onto the hypersurface $\Upsilon$, and
${\cal O}_{...\perp...} \equiv {\cal O}_{...A...}n^A$ for projections onto the normals.  
As an example of this notation, we use ${\cal E}_{ac} \equiv {\cal C}_{a\perp c\perp} =
{\cal C}_{ABCD}h^A_ah^C_cn^Bn^D$ for the `electric' part of the Weyl tensor.
Contracting each of these equations once results in
\be
{\cal R}_{bd} - {\cal R}_{\perp b\perp d} =  R_{bd} - (KK_{bd} - {K_{b}}^{c}K_{cd}),
\label{Gaussc}
\ee
\be
{\cal G}_{a\perp} = {\cal R}_{a\perp} = -(D_b{K^b}_a - D_aK),
\label{Codazzic}
\ee
\be
{\cal R}_{\perp\perp} =  \frac{1}{\alpha}(\delta K - D^2\alpha) - K\circ K, 
\label{Riccic}
\ee
where $K \circ K \equiv K_{ij}K^{ij}$.  
Contracting the Gauss equation a second time gives
\be
-2{\cal G}_{\perp\perp} = {\cal R} - 2{\cal R}_{\perp\perp}  = R - K^2 + K\circ K
\label{Gausscc}.
\ee
The above equations are also related to the projections of the Einstein tensor ${\cal G}_{AB}$ as shown.

Now to obtain 4-d BEFE, 
one begins by constructing the 4-d Einstein tensor from the contracted Gauss equation 
(\ref{Gaussc}) and the doubly-contracted Gauss equation (\ref{Gausscc}):
\be
G_{ab} = {\cal G}_{ab}  - {\cal R}_{\perp a \perp b} +{\cal R}_{\perp\perp}h_{ab} 
+{K} {K}_{ab} + { {K}_{a}} ^{c}{K}_{bc} 
-\frac{{K}^2 - {K} \circ {K}}{2}h_{ab}.
\label{step1}
\ee 
SMS now take three steps in order to derive their formulation of the BEFE, 
and a fourth step is then often used in practice.  

\noindent {\bf Step I:} Using the definition of the Weyl tensor, 
${\cal R}_{\perp a \perp b }$ is replaced by the `electric' part of the Weyl tensor
${\cal E}_{ab} \equiv {\cal C}_{\perp a \perp b} $ and extra terms built 
from the projections of ${\cal R}_{AB}$.

\noindent {\bf Step II:} The 5-d EFE are 
then assumed, which permits one 
to exchange all 
remaining projections of ${\cal R}_{AB}$ for 5-d  
energy-momentum terms. Only when this is carried out does 
(\ref{step1}) become a system of field equations rather than of geometrical 
identities. 
We refer to such field equations as timelike 
hypersurface EFE (THEFE) since at first sight 
they resemble the 4-d EFE.  
Together with the 5 Gauss and Codazzi constraints that one 
forms from 
using the EFE on (\ref{Gausscc}) and (\ref{Codazzic}), 
the 10 THEFE form a system of 15 equations in place of the 15 5-d EFE.  
The 5 constraints may be seen as consistency conditions on the hypersurface. 
For zero 5-d energy-momentum components ${\cal T}_{a\perp}$, 
the last 4 of these correspond to 4-d energy-momentum conservation on the brane. 

\noindent {\bf Step III:} To derive braneworld EFE, which are a special subcase of THEFE, one uses 
normal coordinates and chooses the braneworld energy-momentum tensor ansatz 
\be
{\cal T}_{AB} = {Y}_{AB}\delta(z) - \Lambda g_{AB} \mbox{ } , \mbox{ } \mbox{ } 
{Y}_{AB} \equiv (T_{AB} - \lambda h_{AB} ) \mbox{ } , \mbox{ } \mbox{ } T_{AB}n^A = 0 
\label{BWEM}
\ee
where $T_{ab}$ is the energy-momentum of the matter confined to the brane.   $\Lambda$ and $\lambda$ are 5-d and 4-d 
cosmological constants respectively.
One then adopts junction conditions (JC) to hold across the brane, in particular
\be
[{K}_{ab}]_-^+ \equiv K_{ab}^+ - K_{ab}^- = \kappa_5^2 
\left(
{Y}_{ab} - \frac{{Y}}{3}h_{ab}
\right),
\label{prez2}
\ee
(where we make explicit a $\kappa_5^2$ proportional to the 5-d gravitational constant).   
With the additional supposition of $Z_2$ symmetry\fn{The difference 
in the sign of (\ref{prez2}) between this letter and SMS's paper is due 
to our use of the opposite sign convention in the definition of extrinsic curvature.
We compensate for this in subsequent formulae by also defining $K_{ab} = -K_{ab}^+$ rather than $+K_{ab}^+$.}
i.e $- K_{ab} \equiv K_{ab}^+ = - K_{ab}^- $   
\be
{K}_{ab} = - \frac{\kappa_5^2}{2}\left({Y}_{ab} - \frac{{Y}}{3}h_{ab}\right) = 
- \frac{\kappa_5^2}{2}\left(T_{ab} - \frac{{T - \lambda}}{3}h_{ab}\right) 
\label{41jc}
\ee
as derived in the next section.  
The SMS braneworld EFE then read
\be
G_{ab} = L^{\mbox{\scriptsize SMS\normalsize}}_{ab}(T) + Q^{\mbox{\scriptsize SMS\normalsize}}_{ab}(T) - {\cal E}_{ab} 
\label{BEFEstr},
\ee
where $Q_{ab}(T)$ is the quadratic term and $L_{ab}(T)$ is the linear (together with zeroth order) term in $T_{ab}$ respectively, 
given by
\be
Q^{\mbox{\scriptsize SMS\normalsize}}_{ab} = \kappa_5^4
\left[ 
\frac{T}{12}T_{ab} - \frac{1}{4}T_{ad}{T^{d}}_{b} + 
\left(
\frac{T \circ T}{8} - \frac{T^2}{24}
\right)
h_{ab}
\right], 
\label{QBEFE}
\ee
\be
L_{ab}^{\mbox{\scriptsize SMS\normalsize}} = -\frac{\kappa_5^2}{2}\left(\Lambda + \frac{\kappa_5^2}{6}\lambda^2\right)h_{ab} + 
\frac{\kappa^4_5}{6}\lambda T_{ab}
\label{LBEFE}.
\ee

\noindent {\bf Step IV:} In contrast to the 4-d EFE, the SMS braneworld EFE 
are not closed since ${\cal E}_{ab}$ is unspecified. 
To close the system, SMS write down further equations (which give a large 
third-order brane-bulk system) for the `evolution' away from the timelike 
brane of the `electric' and `magnetic' parts of the manifestly 5-d Weyl 
tensor \cite{SMS}.  Given the complexity of this full brane-bulk third-order system, 
other workers have often treated the SMS braneworld EFE alone.  
This involves the ad hoc prescription of the functional form of ${\cal E}_{ab}$.  
This is sometimes completely thrown away (see e.g \cite{pink}).  
It is sometimes decomposed according to a standard procedure \cite{Maartensdec}.  
Because the original functional form is unknown, the functional 
form of each of the parts defined by the decomposition is also unknown.\fn{The 
unknowns are sometimes kept, e.g as the `Weyl charge' for black holes in \cite{BH}.}   
Some parts are then set to zero and others are restricted by unjustified but 
convenient ans\"{a}tze.  In particular the anisotropic stress part ${\cal P}_{ab}$ 
is sometimes set to 0 (see e.g \cite{green}) or otherwise restricted \cite{yellow}.  
The radiative perfect fluid part is often kept, but is then argued to be small 
(despite containing an unknown factor) in the 
in the inflationary 
\cite{MWBH} and perturbative \cite{Maartensdec, cpert} treatments. 
Having dealt with the Weyl term in such a way, 
the above form of 
$Q_{ab}^{\mbox{\scriptsize SMS\normalsize}}$ is then
often taken to be uniquely defined and is the starting-point of many works in brane cosmology.  
  
As we shall see in the next section, the split into a term which `characterizes the bulk' 
$- {\cal E}_{ab} \equiv B_{ab}^{\mbox{\scriptsize SMS\normalsize}}$ and a term `on the brane'   
$L^{\mbox{\scriptsize SMS\normalsize}}_{ab}(T) + Q^{\mbox{\scriptsize SMS\normalsize}}_{ab}(T)$ 
is in fact highly non-unique.  

\section{Non-uniqueness of split of BEFE terms into `brane' and `bulk' terms}

The Weyl term in the SMS braneworld EFE (\ref{BEFEstr}) has been the subject of much mystery.    
Our aim here is not to argue about what form the Weyl term may take 
in particular solutions of the system (e.g zero everywhere).  Rather 
we show that different formulations of the full brane-bulk system exist which - although 
completely equivalent - lead to very different splits of the non-Einsteinian terms in the 
BEFE into `brane' and `bulk' terms.  Some of these reformulations have no explicit Weyl term 
present in the BEFE.  

To proceed we first note that the issue of the presence of a Weyl term in systems 
derived from splitting the Einstein equations w.r.t non-null hypersurfaces 
is an old one, which has nothing to do with the brane 
energy-momentum ansatz, since the Weyl term is present
from the start in (\ref{step1}). 
At the relevant level, it also has nothing to do with 
the signature and dimension of the spacetime and the signature of the codimension 1 hypersurface 
(an example of which is the brane) w.r.t which the split is performed.   
Thus as discussed in \cite{AT} the presence of such a Weyl term has been considered 
in the development of the GR Cauchy problem.  As we recollect below, exactly the same 
procedure is used in the derivation of the junction conditions \cite{jns}.  
\it The Weyl term does not occur if one chooses a formulation in which it is removed early on 
by use of the Ricci equation\normalfont. If this is not done, as is the case in the
SMS formulation, then
one requires subsequent use of Bianchi identities and the Ricci equation 
in order to close the system at third order.  

We next provide the derivation of the junction conditions to illustrate that the suggested use 
of the Ricci equation is entirely natural. This derivation also demonstrates some of the 
different ways in which the SMS construction of the BEFE is non-unique.  
While such formulations are equivalent to SMS's, each has a distinct split into `brane' and 
`bulk' terms.  We will then systematically list and explain the sources of non-uniqueness.     

In embedding a 4-d timelike thin matter sheet in a 5-d spacetime manifold,   
whereas to have well-defined geometry one requires the metric 
to be continuous across the thin matter sheet yielding the
JC
\be
[h_{ab}]^+_- = 0,
\label{jcf}
\ee
discontinuities in certain derivatives of the metric are permissible.
Consider the 3 projections of ${\cal G}_{AB}$.  One uses 
(\ref{Gausscc}) and (\ref{Codazzic}) to obtain 
${\cal G}_{a \perp}$ and ${\cal G}_{\perp\perp}$
and for 
${\cal G}_{ab}$ one proceeds as SMS do by forming 
(\ref{step1}) and then applying the following steps:

\noindent
{\bf Step V:} The Ricci equation (\ref{Riccieq}) is used to remove all the ${\cal R}_{\perp a\perp b}$. 

\noindent
{\bf Step VI:} The contracted Ricci equation (\ref{Riccic}) is used to remove all the ${\cal R}_{\perp\perp}$.  Thus one arrives at 
\be
G_{ab} = {\cal G}_{ab}
-\left(
\frac{
\delta {K}_{ab} -  D_{a} D_{b}\alpha
}
{\alpha} + {{K}_{a}}^{d} {K}_{bd}
\right)
\\
+\left(
\frac{\delta {K} -  D^2\alpha}{\alpha} - {K} \circ {K}
\right)
h_{ab}
+{K} {K}_{bd} - { {K}_{a}} ^{d}{K}_{bd}
-\frac{{K}^2 - {K} \circ {K}}{2}h_{bd}.
\ee
Then passing to normal coordinates and rearranging via the definition of extrinsic 
curvature (\ref{ecd}) to form the completed normal derivative
$\frac{\pa}{\pa z}({K}_{ab} - {h}_{ab}{K})$, one obtains the following geometrical identity:
\be
G_{ab} = {\cal G}_{ab}  -\left( {K} {K}_{ab} + 2{ {K}_{a}} ^{d}{K}_{bd}
                                                                     + \frac{{K}^2 + {K} \circ {K}}{2}h_{ab} \right)
                                                                     - \frac{\pa}{\pa z}\left({K}_{ab} -  h_{ab}{K} \right).
\label{Israel}
\ee
$$\mbox{Now performing} \mbox{ } \begin{array}{c} \mbox{lim} \\ \epsilon \longrightarrow 0\end{array} \int_{-\epsilon}^{+\epsilon}{\cal G}_{AB}dz \mbox{ }
\mbox{one obtains the JC}
\mbox{ } \mbox{ }\mbox{ } \mbox{ }\mbox{ } \mbox{ }\mbox{ } \mbox{ }\mbox{ } \mbox{ }\mbox{ } \mbox{ }\mbox{ } \mbox{ }\mbox{ } \mbox{ }
\mbox{ } \mbox{ }\mbox{ } \mbox{ }\mbox{ } \mbox{ }\mbox{ } \mbox{ }\mbox{ } \mbox{ }\mbox{ } \mbox{ }\mbox{ } \mbox{ }\mbox{ } \mbox{ }
\mbox{ } \mbox{ }\mbox{ } \mbox{ }\mbox{ } \mbox{ }\mbox{ } \mbox{ }\mbox{ } \mbox{ }\mbox{ } \mbox{ }\mbox{ } \mbox{ }\mbox{ } \mbox{ }
\mbox{ } \mbox{ }\mbox{ } \mbox{ }\mbox{ } \mbox{ }\mbox{ } \mbox{ }\mbox{ } \mbox{ }\mbox{ } \mbox{ }\mbox{ } \mbox{ }\mbox{ } \mbox{ }
\mbox{ } \mbox{ }\mbox{ } \mbox{ }\mbox{ } \mbox{ }\mbox{ } \mbox{ }\mbox{ } \mbox{ }\mbox{ } \mbox{ }\mbox{ } \mbox{ }\mbox{ } \mbox{ } $$
\be
[{\cal G}_{\perp\perp}]_-^+ = 0 \mbox{ } , \mbox{ } \mbox{ } [{\cal G}_{a\perp}]_-^+ = 0 \mbox{ },
\label{jc0}
\ee
\be
[{\cal G}_{ab}]_-^+ =  [{K}_{ab} - K h_{ab}]^+_-.
\label{primjn}
\ee

\noindent If one now uses the 5-d EFE (Step II),
with thin matter sheet energy-momentum
\be
{Y}_{AB} =
\begin{array}{c}
\mbox{lim} \\
\epsilon \longrightarrow 0
\end{array}
 \int_{-\epsilon}^{+\epsilon} {\cal T}_{AB} dz,
\ee
one obtains the JC 
\be
0 = {Y}_{\perp\perp} \mbox{ } , \mbox{ } \mbox{ } 0 = {Y}_{a\perp} \mbox{ },
\label{jc0M}
\ee
from (\ref{jc0}) and additionally, via a trace reversal, 
the JC (\ref{prez2}) from (\ref{primjn}).

Clearly this is a different choice of procedure from that used by SMS to formulate their THEFE.
This illustrates that one has a number of 
choices as to what geometrical objects are
present in the THEFE, and hence in the BEFE.
The terms in the BEFE that require the knowledge of the bulk constitute
the `bulk' part of the bulk-brane split.   
The terms in question are higher-dimensional curvature terms, in particular those 
terms (such as the Weyl term in SMS's formulation) which cannot be replaced via the 5-d EFE,   
and normal derivatives of objects such as the extrinsic curvature [which occur in
Eq. (\ref{Israel})]. The key point is that we can change around which of these 
objects occur in the THEFE by using well-known geometrical identities 
and doing so changes the form of the term quadratic 
in $K_{ab}$, and hence the term quadratic in $T_{ab}$ 
when one passes to the BEFE by adopting the 
braneworld energy-momentum tensor ansatz (\ref{BWEM}).  
For example, using the set of steps at the start of this 
section leading to Eq. (\ref{Israel}),
the EFE and the braneworld 
energy--momentum ansatz lead to BEFE of the following form
\be
G_{ab} = L_{ab} + Q_{ab} + B_{ab},
\label{ex0}
\ee
with
\be
Q_{ab} = -\frac{\kappa_5^4}{72}
\left[
36{T_{a}}^{d}T_{bd} - 15TT_{ab} + (9T\circ T + 5T^2)h_{ab}
\right],
\label{ex0q}
\ee
\be
L_{ab} = -\frac{\kappa_5^4}{9}[12T_{ab}\lambda - (7T \lambda  - 8\lambda^2)h_{ab}]   
+ \kappa_5^2[T_{ab} - (\lambda + \Lambda) h_{ab}]
\label{ex0l},
\ee
\be
B_{ab} = \frac{\pa}{\pa z}
\left(
Kh_{ab} - K_{ab}
\right),  
\label{ex0b}
\ee
where $L_{ab}$ are the terms linear in the brane energy-momentum 
and $B_{ab}$ is the `bulk' part of the split. 
It is now clear that the split into $B_{ab}$ and $L_{ab} + Q_{ab}$ is non-unique, 
since the above split is clearly not the same as that due to SMS.  Rather, the outcome 
of the split depends on which geometrical objects are used in the formulation.

With what has been clarified by the above recollection of the derivation of the JC in mind, 
we now provide a comprehensive list of steps that may in general be applied in the 
construction of BEFE.

\noindent Steps I and II together mean that the Weyl `bulk' term ${\cal E}_{ab}$ is equivalent to 
a Riemann `bulk' term together with matter terms.  
This swap by itself involves no terms which are 
quadratic in the extrinsic curvature.

\noindent Step V says that the Riemann `bulk' term is equivalent 
to a `bulk' term containing of hypersurface derivative of the extrinsic 
curvature together with a ${K}_{ad}{{K}^{d}}_{b}$ term.
\noindent Steps VI and II together say that the hypersurface derivative  
of the trace of the extrinsic curvature is equivalent to a matter term together with a ${K} \circ {K}$ term.  
\noindent The idea is to use Steps VI and II, and Step V, on
arbitrary proportions (parametrized by $\mu$ and $\nu$)
of ${\cal R}_{\perp a\perp b}$ and of ${\cal R}_{\perp\perp}$:

\be
\begin{array}{ll}
G_{ab} & = {\cal G}_{ab} - (1 +\nu){\cal R}_{\perp a\perp b}\mbox + (1 - \mu){\cal R}_{\perp\perp}h_{ab}  
+\frac{1}{\alpha}
\left[
\nu(\delta {K}_{ab} -  D_{a} D_{b}\alpha)  
+ \mu(\delta {K} -  D^2\alpha)h_{ab}
\right] \\&
+ 
\left(
{K} {K}_{ab} + (\nu - 1) {{K}_{a}}^{d} {K}_{bd} 
- \frac{{K}^2}{2}h_{ab}  +  (\frac{1}{2} - \mu) {K} \circ {K}h_{ab}
\right).
\end{array} 
\label{2param}
\ee

\noindent {\bf Step VII:} We are furthermore free to choose to characterize 
the bulk in terms of hypersurface derivatives $\delta$ (which are normal derivatives $\frac{\pa}{\pa z}$ 
in normal coordinates) of objects which may be 
related to the extrinsic curvature by use of the metric tensor. Relating the 
hypersurface derivatives of these new objects to those of the 
extrinsic curvature requires taking metrics inside or outside the hypersurface derivative.  
Examples of such moves, which all follow from the product rule 
and the definition of extrinsic curvature, are \bf(i)\normalfont
\be
\delta K = h^{ab} \delta{K}_{ab} + 2\alpha {K} \circ {K}.
\label{63}
\ee

\noindent {\bf (ii)} We could choose to work with objects with raised indices e.g. ${K^{a}}_b$:
\be
\delta{K_{ab}} = \delta (h_{ac}{K^{c}}_{b}) = h_{ac}\delta {K^c}_a - 2\alpha K_{ac}{K^c}_b.
\ee

\noindent {\bf (iii)} We could define an object by removing a portion $\eta$ of the trace from the extrinsic curvature: 

\noindent${K}^{\mbox{\scriptsize $\eta$\normalsize}}_{ab} \equiv {K}_{ab} - \eta{K}h_{ab}$. Then
\be
\delta{K}_{ab} = \delta{K}^{\mbox{\scriptsize $\eta$\normalsize}}_{ab} + \eta
\left(\delta{K}h_{ab} - 2\alpha{K}{K}_{ab}
\right).
\label{64}
\ee    

\noindent {\bf Step VIII:} We are also free to choose to characterize the bulk 
in terms of the normal derivatives of densitized objects such as $h^{\xi_1}{K}_{ab}$ 
and $h^{\xi_2}{K}$ for $h = \mbox{det}h_{ab}$.  
Then the identities relating the hypersurface derivatives of these objects to those of $K_{ab}$ are
\be
\delta{K}_{ab} = h^{-\xi_1}\delta(h^{\xi_1}{K}_{ab}) + 2\alpha\xi_1{K}{K}_{ab},
\label{65}
\ee  
\be
\delta{K} = h^{-\xi_2}\delta(h^{\xi_2}{K}) + 2\alpha\xi_2{K}^2. 
\label{66}
\ee

To illustrate that trace-removed and densitized objects are entirely natural, 
we may recall (out of many examples in the literature \cite{AT}) the 
`gravitational momenta'  $p_{ab} = -\sqrt{h}({K}_{ab} - {K}h_{ab})$.
Also the move whereby $h_{ab}$ is taken inside the derivative to 
form a complete normal derivative in Eq. (\ref{Israel})
may be interpreted as the $\eta =1$ case of (iii).

We note from this thorough consideration of possible `bulk' terms 
that all 4 `brane' terms quadratic in the extrinsic curvature in the THEFE can be changed 
independently. There are thus many bulk characterizations such that all 
4 of these terms, and hence $Q_{ab}(T)$, are zero.  
In the next section we illustrate this point with the help
of two examples.  

\section{Examples of BEFE with No Quadratic Terms}

The diversity of splits into `bulk' and `brane' terms ensures that the truncation Step IV, 
whereby whichever bulk term is present is neglected, produces all possible combinations 
of quadratic terms depending on the choice of split employed.   
Here we illustrate by simple examples that were any such truncation used, 
then which particular truncation it is 
could lead to big differences in the remaining `braneworld physics'.    
We do this by building splits into `bulk' and `brane' terms in which 
no $Q_{ab}$ at all is left in the `brane' term. 
Thus for example rather than brane FLRW cosmology (which includes $\rho^2$ terms)
\cite{Bin2, MWBH, Maartensdec} 
one would obtain standard FLRW cosmology (with a $\rho$ term alone).  

As our first example, we take as the primary object the antidensitized extrinsic curvature
$\underline{K}_{ab} \equiv \frac{   K_{ab}   }{   \sqrt{ h  }   }$
so that the bulk is (in part) characterized by
its normal derivatives.
Thus, using (\ref{2param}), (\ref{63}), (\ref{65}) and
the 5-d Einstein field equations, we obtain 
\be
G_{ab} = {L}_{ab} + B_{ab},
\ee
with
\be
L_{ab} = \frac{{\cal T}_{ab}}{3} + \frac{1}{6}\left(5{\cal T}_{\perp\perp} - {\cal T}\right)h_{ab} \mbox{ } , \mbox{ } \mbox{ }
B_{ab} = -2{\cal E}_{ab} + \sqrt{h}
\left(
\frac{\pa {\underline{K}_{ab}}}{\pa z}
- \frac{1}{2}h^{cd}
\frac{\pa \underline{K}_{cd}}{\pa z} h_{ab}
\right).
\ee
This example may be reformulated, using (\ref{63}), (\ref{64}) and (\ref{65}),
so that the primary objects
are the `gravitational momenta', in which case 
\be
B_{ab} = -2{\cal E}_{ab} - \frac{1}{\sqrt{h}}
\left[
\frac{\pa p_{ab}}{\pa z} +
\left(
\frac{1}{3}\frac{\pa p}{\pa z} - \frac{1}{2} h^{cd}
\frac{\pa p_{cd}}{\pa z}
\right)h_{ab}
\right].
\ee
Truncating these equations by neglecting the bulk term $B_{ab} $,
and assuming perfect fluid matter on the brane with equation of state $P = (\gamma - 1)\rho$, we obtain the corresponding
braneworld Friedmann equation
\be
\frac{3}{a^2}(\dot{a}^2 + k) = 
\kappa_5^2\left[\rho\left(-\frac{1}{3} + \frac{\gamma}{2}\right) + \frac{\Lambda - \lambda}{3}\right],
\ee
which clearly does not possess a quadratic term in $\rho$.

One may consider the BEFE formulation in Eqs (24-27) 
as having a `bulk' term which contains no ${\cal E}_{ab}$ at all. 
We then ask if it is possible to find a formulation in which neither $Q_{ab}(T)$ nor 
${\cal E}_{ab}$ feature. With these restrictions in mind, we found the following example. 
By considering as our primary object the densitized extrinsic curvature with one index raised 
${\overline K}^a\mbox{}_b \equiv \sqrt{h}{K^a}_b$, 
we found the corresponding BEFE to have a `bulk' term composed entirely of normal derivatives:
\be
G_{ab} = L_{ab} + B_{ab},
\label{ex2}
\ee
\be
L_{ab} = {\cal T}_{ab} + \frac{1}{2}\left({\cal T}_{\perp\perp} - \frac{\cal T}{3}\right)h_{ab} 
\mbox{ } , \mbox{ } \mbox{ }
B_{ab} = 
\frac{1}{\sqrt h}\left( \frac{1}{2}\frac{\pa \overline{K}}{\pa z} h_{ab} - \frac{\pa {\overline K}^c\mbox{}_b}{\pa z} h_{cb} \right).
\label{ex2lb}
\ee
Truncating this and using $P = (\gamma - 1)\rho$, the corresponding braneworld 
Friedmann equation is
\be
\frac{3}{a^2}(\dot{a}^2 + k) = 
\kappa_5^2\left[\rho\left(\frac{1}{3} + \frac{\gamma}{2}\right) + \frac{2\Lambda + \lambda}{3} \right],
\ee
which again is devoid of a $\rho^2$ term.

The explicit absence of quadratic terms in the above examples 
is due to choosing variables in which the quadratic terms have been 
entirely incorporated into derivatives off the hypersurface, knowledge of which 
one would assume would require unavailable knowledge about the bulk.  
Thus the quadratic terms are implicitly present in the full systems above, as they must be since these 
systems are equivalent to that of SMS.  However, were one to truncate the `bulk-like' terms in these 
formulations, one would find that one had inadvertently thrown away the `brane-like' quadratic terms 
as well!  We emphasize that we are not advocating that any of these Friedmann equations arising by 
truncation of bulk terms should be taken seriously.  On the contrary, the aim of 
these examples is to demonstrate with simple calculations (rather than involving perturbative methods or anisotropic models)  
the fact that in general truncations result in inequivalent residual `braneworld physics'.   Thus truncations should be avoided 
in the study of the SMS braneworld.

It is important to note here that in the
case of the SMS formulation with AdS bulk (which has ${\cal E}_{ab} =0$),
the full and the truncated systems coincide. In this case
SMS's formulation (together with its $\rho^2$ term) is
equivalent to all the full formulations. However it is not typical for the bulk to have ${\cal E}_{ab} =0$, 
which means that this convenient adaptation is of limited use.  
By the very same argument, we caution that perhaps some 5-d spacetimes possess embedded 
$B_{ab} = 0$ hypersurfaces for some $B_{ab} \neq {\cal E}_{ab}$ which would amount to 
the full brane-bulk system admitting solutions containing branes with a non-SMS quadratic term.  
Furthermore, were this to occur for some $B_{ab}$ corresponding to no BEFE quadratic terms, then  
this would amount to the full brane-bulk system admitting solutions containing branes with no 
associated quadratic term.  Since each 5-d spacetime contains an infinity of embedded 4-d timelike 
hypersurfaces for which any of the $B_{ab} = 0$ conditions might hold, proving or disproving the 
above possibilities is a difficult geometrical problem.

\section{Discussion}

Whereas the particular SMS formulation (often just the BEFE with ${\cal E}_{ab}$ 
or ${\cal P}_{ab}$ thrown away)  
has often been taken to be the starting point for GR-based brane cosmology, we have shown 
that there are many choices of formulation of braneworlds by use of geometrical identities.  
Whereas these formulations are clearly equivalent for the full brane-bulk system, in each case 
different BEFE terms have a manifest `bulk' origin because the geometrical identities used mix 
up `bulk' and `brane' terms.  Then were one to throw away the `bulk' term in each case,  
one would obtain inequivalent truncated systems.  It is important to bear in mind that one 
does not \it a priori \normalfont know whether in general 
it is more or less dangerous to throw away one type of $B_{ab}$ (e.g ${\cal E}_{ab}$) than any other.  
In SMS's formulation, the quadratic term is expressed in terms of the energy-momentum residing on the 
brane, whereas the Weyl term is a portion of a higher-dimensional tensor, so one might feel justified in 
throwing away the one but not the other.  But in other formulations both of these terms are replaced 
by combinations of other quadratic terms and decidedly bulk-like derivatives off the hypersurface.  
Indeed we have shown that there exist formulations in which both of these terms are replaced entirely by 
derivatives off the hypersurface.  Thus our point is that there is no clear concept of which truncations 
are or are not dangerous.  Furthermore, such truncations lead to inequivalent residual `braneworld physics' 
as exemplified by the braneworld Friedmann equations with no $\rho^2$ term in them.  
Thus we have an argument against 
performing {\it any} truncations at all, and that includes throwing away ${\cal E}_{ab}$ or 
${\cal P}_{ab}$. 
Our examples serve as a warning that to understand the SMS braneworld, one must consider the full 
brane-bulk system.  

Finally, having argued in favour of the study of the full brane-bulk system, we note that the availability 
of formulations pointed out in this letter may greatly facilitate this study.
SMS's formulation of the full brane-bulk system is third-order (in the metric) since 
it includes evolution equations for the `electric' and `magnetic' parts of the Weyl tensor.  
Our point is that this system has this form only because the Ricci equation (\ref{Riccieq})  
is not used early on in deriving the formalism.  Thus the `electric' Weyl term remains 
within the BEFE's as an extra unknown, and Bianchi identities together with the Ricci equation are required  
to evolve it [which in turn involves the `magnetic' Weyl term 
as yet another unknown].   We suggest that before a detailed study of 
the brane-bulk system in this particular third-order formulation of SMS is carried out, 
it is well worth investigating the reformulations which can be obtained along the lines of this letter.  
From the point of view of PDE theory, knowing precisely which reformulations are available 
for a given system is of central importance toward providing theorems.     
In particular here, some formulations of the brane-bulk systems are closed at second order due to the early use 
of the Ricci equation (\ref{Riccieq}) in deriving these formulations.   These second-order formulations 
include our BEFE (\ref{ex0}--\ref{ex0b}), the GR Cauchy problem analogue in \cite{AT} and our second example 
(\ref{ex2}, \ref{ex2lb}).  We emphasize that each such formulation consists of just 15 at most second-order 
equations (the BEFE together with the Gauss--Codazzi constraints) rather than the much larger number of mostly 
third-order equations in SMS's formulation.

\vskip 0.1in

\noindent\large \bf
Acknowledgments
\normalfont \normalsize
\noindent EA is supported by PPARC.  
We thank Naresh Dadhich, Henk van Elst, James Lidsey, Roy Maartens and Malcolm MacCallum for discussions.

\end{document}